# From the scale model of the sky to the armillary sphere

**Alejandro Gangui**
IAFE/Conicet and Universidad de Buenos Aires, Argentina.

**Roberto Casazza**
Universidad Nacional de Rosario and Universidad de Buenos Aires, Argentina.

**Carlos Paez**
Instituto Superior de Formación Docente N° 29, Merlo, Buenos Aires, Argentina.

*It is customary to employ a semi-spherical scale model to describe the apparent path of the Sun across the sky, whether it be its diurnal motion or its variation throughout the year. A flat surface and three bent semi-rigid wires (representing the three solar arcs during solstices and equinoxes) will do the job. On the other hand, since very early times, there have been famous armillary spheres built and employed by the most outstanding astronomers for the description of the celestial movements. In those instruments, many of them now considered true works of art, Earth lies in the center of the cosmos and the observer looks at the whole "from the outside". Of course, both devices, the scale model of the sky and the armillary sphere, serve to represent the movement of the Sun, and in this paper we propose to show their equivalence by a simple construction. Knowing the basics underlying the operation of the armillary sphere will give us confidence to use it as a teaching resource in school.*

**An ancient astronomical instrument**

In Latin, *armilla* means ring or bracelet. From this comes the name armillary sphere, used to designate a model or three-dimensional representation of the celestial sphere, the apparent path of the Sun in the sky and some significant astronomical elements such as the celestial equator, tropics, polar circles and the ecliptic. The gadget consists of a set of concentric rings arranged around a small sphere which serves as Earth. Dating back of old, its invention is attributed, among others, to Eratosthenes of Cyrene (276-195), Greek astronomer, mathematician and geographer who lived mainly in Alexandria, and whose famous library he directed.

The old armillary spheres are extremely ingenious mechanisms and constitute very attractive objects. Presenting them in a school context allows us to exploit both features, but has in addition a most important virtue for the work of the teacher: it helps to guide the students through the historical path of science [1]. It is, for example, a valuable resource for assessing the relevance of the Ptolemaic (or geocentric) view of the cosmos before the rise of heliocentrism. However, in their Astronomy courses, teachers in general prefer to employ a simple scale model made of wires to represent the Sun's apparent diurnal motion (Figure 1), and rarely use an armillary sphere.

Figure 1 shows the "wire" model that represents the apparent path of the Sun across the sky in three positions representing four particular days of the year. The observer is located at the center of the horizontal base. The sphere impaled on one of the wires represents the Sun and can be moved along the wire to simulate its position in the sky at different times of the day: when it reaches the highest point it indicates the solar noon. For the northern hemisphere (and for a site located north of the Tropic of Cancer), one can assume that the semicircular wire arcs correspond (from left to right) to the winter solstice, the two equinoxes and the summer solstice. In this case, the wire model is oriented with the south cardinal point to the left of the picture (in fact, in the northern hemisphere, the solar diurnal arcs are inclined towards the south). The separation of the attachment points of the wires at the base depends on the latitude, and was already discussed qualitatively in Reference [2].

**A simplified armillary sphere for the classroom**

In order to work in class with the armillary sphere it is advisable to use a simplified model, like the one shown in Figure 2. It consists of a small ball which represents Earth, and a set of metal rings that surround it. The rotation axis of the Earth serves to maintain the whole set together and is screwed to the largest of the rings. The Earth's axis is up about 35° from the horizontal plane, given by the table or surface. This corresponds, for example, to the latitude of the city of Albuquerque and can be varied to represent the movement of the sky at any terrestrial latitude.

The largest ring corresponds to the meridian of the location for which the sphere is calibrated and defines a vertical plane passing through the two celestial poles. The moment in which the Sun moves from one side to the other of this plane is called solar noon. That's why we talk of hours AM (ante meridiem) and PM (post meridiem), to designate the morning and afternoon respectively.

With the armillary sphere set as shown in the pictures, the top of the Earth's axis points to the north celestial pole (very close to the star *Polaris*), which in the latitude of Albuquerque is located at an altitude of about 35° above of the north cardinal point.

Consider now the two smaller rings, tilted approximately 23.5° among themselves, that we can imagine as two circles projected on the dome of the sky. The ring defining a plane exactly perpendicular to the axis of the world is the celestial equator; note that it exactly cuts the horizon at the cardinal points east and west. The other is the ecliptic. As we will see, our armillary sphere allows us to represent the apparent movement of the Sun in the sky, both during the day and during the year, for a terrestrial observer considered fixed and motionless in the center of the cosmos (the center of the instrument). It thus provides us with the classical geocentric representation of the world, like the wire model of Figure 1.

**An attractive teaching resource**

Take the ecliptic ring. The location of the Sun in it, for one day, is very close to its location for the following day: the difference is of the order of one degree, as the full turn of the Sun across the backdrop of the stars (360 degrees) occurs in approximately one year (about 365 days). Day after day, as the Sun moves along the ecliptic, its separation from the equatorial ring also changes, since both rings are not parallel. This departure is measured with the equatorial coordinate called "declination".

When the Sun is located to the north of the celestial equator, during spring and summer in the northern hemisphere, its declination is by convention positive, while it becomes negative when, during fall and winter, it moves to the south of the equator. It follows from this that when the Sun is just at the junction of the two rings, i.e. when it is on the celestial equator, its declination is zero. This is what happens twice a year, at the equinoxes. These two days, at solar noon, the Sun is right at the zenith of the inhabitants of Earth's equatorial line.

In Figure 3 we use a small orange ball to represent the Sun. The location of this ball along the ring of the ecliptic was chosen to represent the situation of the summer solstice in the northern hemisphere, when the Sun reaches its farthest point away form the celestial equator, that is, when its declination is maximum and is at 23.5°. Furthermore, the armillary sphere was adjusted to yield the position of the Sun at solar noon.

If we rotate the movable part of the device around the axis of the world (see Figure 2), we see that the Sun ball describes an arc in the sky: it emerges from the eastern horizon, rises, reaches its highest point, then begins to descend and finally hides below the western horizon, as it is shown in the sequence of images of Figure 4. The virtual arc or path thus described by the Sun is similar to that shown by the upper semicircular wire arc (the longest and the one to the right) of the wire model in Figure 1. This in fact corresponds to the diurnal motion of the Sun on the summer solstice. That day, in the northern hemisphere, the Sun emerges from the horizon as far north as possible from cardinal point east, and sets north of west.

If we now adjust the armillary sphere and the Sun ball to represent the winter solstice, when the Sun reaches its maximum negative declination (-23.5°), and we repeat the series of photographs, we get the sequence of Figure 5. For an observer located in the northern hemisphere, in Albuquerque for example, the apparent arc described by the movement of the Sun during that day will be the shortest of the year and it corresponds to the shortest semicircular wire arc (the one to the left) of the wire model of Figure 1.

From what we have explained so far, it follows that when the Sun's location on the ecliptic ring is half way between the solstices, i.e. when its declination is zero and is located in the equinoctial points, its diurnal arc will be represented by the intermediate wire arc of the wire model and also by the celestial equator in the armillary sphere. Indeed, these two days per year, the Sun crosses the celestial equator, which means that those who live along Earth's equatorial line will have it exactly overhead (at the zenith) at noon on these dates. In addition, during the equinoxes, for symmetry reasons and if we ignore our planet's atmosphere, all terrestrial observers will witness nights lasting 12 hours and another 12 hours of daylight.

**Final words**

The construction of both devices presented in this article may represent an enlightening activity for the teacher to carry on with her/his students. We consider that hands-on activities like the one presented here are very welcome and important for astronomy education, where spatial thinking and actual outdoor observations, although key for an adequate instruction, are seldom given enough time. The model with wire solar arcs is certainly something simple to manufacture. It may be a little more difficult to build the simplified model of the armillary sphere, which requires a few flexible iron hoops and some welding skills. However, with the details given and having fully understood the meaning of each of its parts (the metal rings for the meridian of the site, for the celestial equator and the ecliptic, and the axis of the world), the design of this device is not impossible to address.

If we do not have the ability to manipulate metals and welds, we may also attempt the construction of the armillary sphere with bent semi-rigid wires, as we used in the wire model. Another option is to use a bicycle wheel (the bicycle tire and axle will represent the celestial equator and the axis of the world, respectively) to which can be added, tilted, a wire or circular hoop of equal perimeter, which will represent the ecliptic. If a student holds with both hands the wheel axle, tilting it from the horizontal at an angle equal to the latitude of the location of her school, and another student handles the changing position of the Sun ball along the inclined hoop, the Sun's daily path for any day of the year can be easily represented.

Simple elements for exercising the apparent movement of celestial objects in the classroom are never scarce; we just need to understand what we want to represent. In this note we tried to help build up that understanding.

**Acknowledgements**

We thank Prof. María Iglesias for discussions and advice. A.G. acknowledges support from Conicet and from the University of Buenos Aires. R.C. acknowledges support from the National University of Rosario and the University of Buenos Aires.

Alejandro Gangui *is staff researcher at the Institute for Astronomy and Space Physics (IAFE) and is professor of physics at the University of Buenos Aires (UBA).*

Roberto Casazza *teaches Medieval and Renaissance philosophy at the National University of Rosario (UNR) and at the University of Buenos Aires (UBA).*

Carlos Paez *teaches physics at the Teacher Training Institute (ISFD N° 29) in Buenos Aires.*

**Figures**

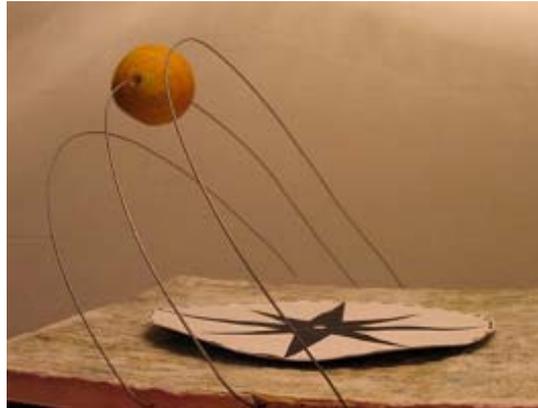

Figure 1: Wire model depicting the path of the Sun in daytime during solstices and equinoxes.

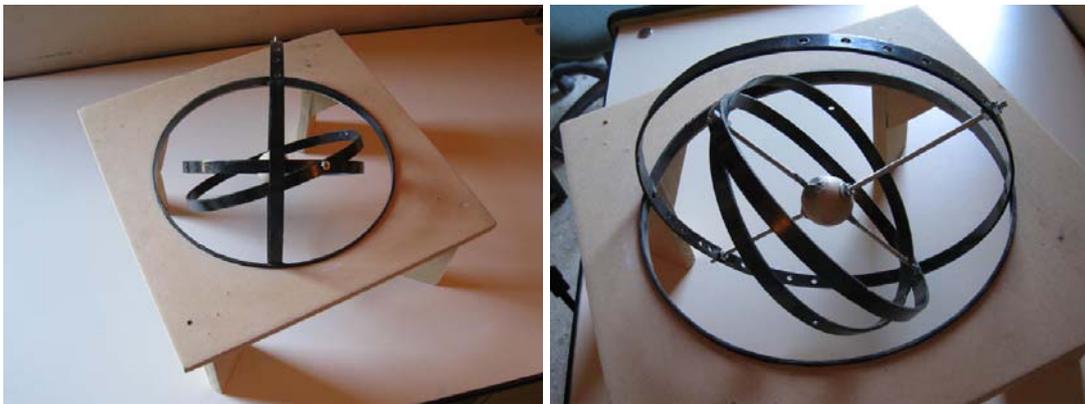

Figure 2: Two views of a simplified model of an armillary sphere. The small ring perpendicular to the axis of the world (or axis of rotation of Earth) represents the celestial equator. The other small ring, inclined with respect to the first one, represents the ecliptic. The Sun is located in a slightly different position of the ecliptic (ring) every day of the year.

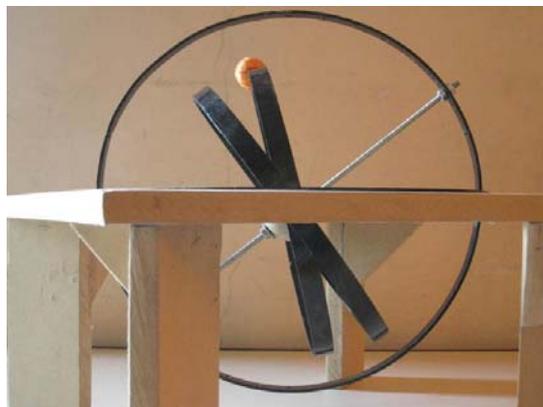

Figure 3: Armillary sphere and a small orange ball representing the Sun, both adjusted to indicate the position of the Sun at solar noon on the summer solstice.

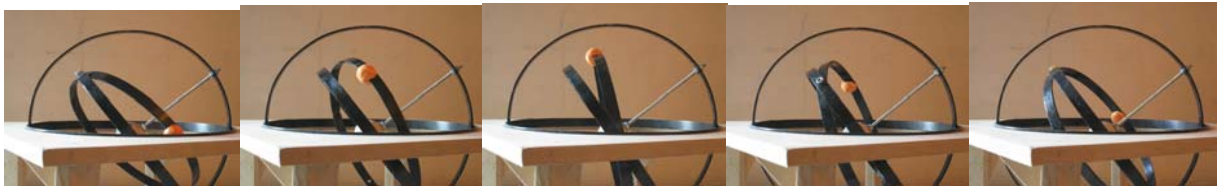

Figure 4: Sequence of photographs of the armillary sphere with a small ball representing the Sun, both adjusted to show the positions of the Sun as seen by a resident of Albuquerque at different times of the day during the summer solstice.

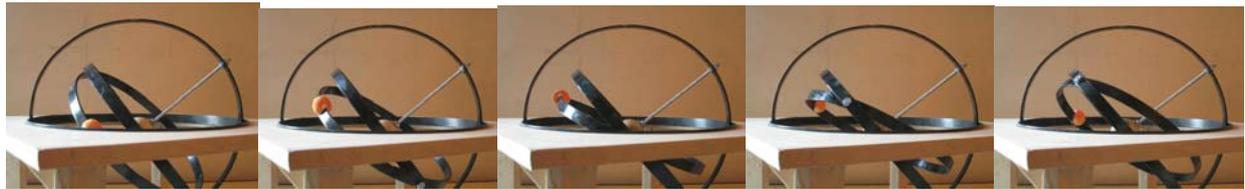

Figure 5: Sequence of photographs of the armillary sphere with the Sun ball, both adjusted to show the positions of the Sun as seen by a resident of Albuquerque at different times of the day during the winter solstice.